# Disruptive Transformation of Enterprise Supply Chain Performance: Strategic Business Networks to Improve Business Value

**Steve Elliot**
University of Sydney Business School, Sydney, Australia
Email: steve.elliot@sydney.edu.au (corresponding author)

**Martin McCann**
The Digital Institute

**Kory Manley**
Ariba, Inc.

## Abstract

In a dynamically changing world, business must transform to survive. Although the necessity for change may be apparent, how to change is not. A useful approach can be to learn how successful pioneers transform a core business function such as procurement: how organisations acquire assets and inputs including facilities, materials and people.

Traditionally, the business objective for procurement has been to increase availability of purchased materials and to reduce costs. Subsequently, the objective became more tactical. Leading procurement practice today is disruptive; beyond cost to creating value. The strategy for market leaders has also transformed; from competition to collaboration. Technology-enabled procurement networks are now key to driving business disruption globally.

Through a rigorous field study of two world-class pioneering corporations, this paper explores how digital disruption is transforming procurement functions. The impact on Enterprise Supply Chains, contributions and implications for current IS theory and practice are discussed.

**Keywords**

Business transformation, supply chain, procurement, business networks, IS theory and practice

## 1　Introduction

In an increasingly volatile world driven by transformative forces including globalization, developments in technology, a changing world order, geo-political instability and the escalating impacts of unsustainable practices, change has become an imperative for business, government and society (NIC 2012, KPMG 2013; Stern 2007; UN SDG 2013; WEF 2010). Although the necessity for transformative change may be apparent; to business executives, why, what, how, where and when to change is not.

The questions, "why, what, where and when" to change may be specific to the organisation. The "how" question, less so. Many, if not most, businesses of any size utilise enterprise systems to provide mission-critical capabilities without which the business would be unable to communicate, organise, manage or perform primary or secondary activities on their value chain (Porter 1985). The role of technology is greater than simply enabling innovation. With globalisation, technology developments have been a major driver of business transformation (Porter and Reinhardt 2007).

Motivations to address the question "why change" include government regulation or directive (European Union 2003a, 2003b), stakeholder influence (Elliot 2013) or societal pressure, as in a 'name and shame' campaign targeting aspects of corporate behaviour (Greenpeace 2007). The remaining questions, "what, where and when" may be a challenge. A useful approach in this situation is to seek to learn from relevant experiences of pioneering businesses. Transformation of a core business function was selected for its potential relevance to organisations as a whole. Procurement is essential for all organisations: business, government and societal. It is how organisations acquire assets and inputs including facilities, materials and people. Traditionally, the business objective for procurement was at an operational level; to acquire the necessary materials and services at the lowest cost. Subsequently, while the business objective remained the same, the focus became more tactical; process improvement to cut inventory and to reduce process costs. Leading procurement practice today is disruptive; moving beyond reducing cost to creating business value. The strategy adopted by market leaders has also transformed; from competing to collaborating. Technology-enabled business networks are now driving business disruption globally (Aberdeen 2011; Hackett 2014).

This paper aims to address business uncertainties, motivate IS researchers and facilitate IS research. First, to assist business executives responding to the drivers of transformative change by exploring how disruption is transforming an Enterprise Supply Chain function. Based on a study of disruptive pioneering business practices a model is proposed to inform the responses of executives uncertain about how to proceed. Second, to stimulate IS researchers to increase the relevance of their work by examining digital disruption of this core business function, and by closing gaps between pioneering business practice and current strategic management and IS theory. Third, to expedite IS research by proposing rigorously-developed models of IS research and practice to be applied, tested and extended. The paper's structure is: a review of relevant literature from three perspectives: reported leading practices, related IS theory and relevant strategic management theory. Research themes and gaps between leading practices and extant theory are identified. The research design, method, and field study of disruptive change in pioneering corporations are described followed by analysis of the research findings, proposal of models, discussion, conclusions and proposals for further study.

## 2    Literature Review

Review of selected literature in the core business function of procurement examines reports describing leading industry practice and compares these reported practices with IS theory. As a whole, IS theoretical contributions are limited. Possibly due to procurement's categorisation as an operational issue focused on cost savings, to date IS researchers have not engaged with this core function as a research priority. Due to the limited IS research addressing pioneering business strategy in this field, literature in strategic business management was explored to identify gaps in current theory. Strategic management literature addressing transformation of the procurement function is also limited. Gaps in the literature for IS and strategic management theory are identified. Note that business reports may use the terms: procurement, sourcing and strategic sourcing, inter-changeably. To avoid confusion in this work, sourcing and strategic sourcing are considered subsets within procurement processes.

### 2.1    Selected themes in literature on leading business practice

Four sources were purposively selected (Miles and Huberman 1994) to identify leading business practice in procurement. A Harvard Business Review study of a corporation describing the process of its procurement transformation in an international retail corporation (Gottfredson et al. 2005); details from a global survey of 315 companies identifying best-in-class procurement performance (Aberdeen 2011) and analysis of the characteristics of companies acknowledged as world-leaders in the procurement field (Hackett 2014). A global survey of 245 companies focusing on Accounts Payable function professionals, identifying the use of IS to automate payment and invoice procurement sub-processes (Ardent 2013).

#### 2.1.1    Procurement transformation process – Business reports

Over the past decades, corporate procurement functions have been evolving from focus at an operational level on availability and purchase price towards procurement as a source of business value through strategic partnerships with selected suppliers (Gottfredson et al., 2005). The transformative process has not been universal nor instantaneous but has been evolving over five distinct stages, see Table 1. A relatively small percentage of corporations have reached the highest, strategic value stage. With few exceptions, notably 7·Eleven's use of creative sourcing partnerships to pioneer entirely new capabilities some 10 years ago (Gottfredson et al. 2005), this small group of businesses has been reluctant to disclose details of their benefits as this presents a source of competitive advantage. Amalgamated responses from business surveys present glimpses of the transformed behaviours, the changing value proposition, the role of procurement and the metrics applied to evaluate the changing activities at each stage (Table 1).

#### 2.1.2    Characteristics of leading practitioners – Business reports

Three key performance criteria can distinguish Best-in-Class performance: spend under the management of the procurement group, procurement contract compliance, and realized / implemented cost savings. Survey results also show that the firms enjoying Best-in-Class performance shared several common characteristics, including higher likelihood of instituting internal collaboration between key internal units and higher likelihood of leveraging an active spend analysis program (Aberdeen 2011). To achieve Best-in-Class performance companies must: leverage e-sourcing solutions to drive higher savings and automate manual strategic sourcing processes; utilize spend analytics to drive visibility into corporate spending and forecast savings for future planning and budgeting; and align overall sourcing activities or processes with the goals and objectives of the organization as a whole. (Aberdeen 2011).

Case examples of leading practice are informative, but rare. The retailer, 7·Eleven, used creative sourcing partnerships to pioneer entirely new capabilities. From being a one stop source convenience store with narrowly focused competitors, it established a consortium to provide multipurpose kiosks in its stores. American Express supplies ATM functions, Western Union handles money wires, and Cash Works furnishes check-cashing capabilities. EDS integrates the technical functions of the kiosks. 7-Eleven maintains control over the data on how customers use the kiosks - which it views as critical to its competitive edge (Gottfredson et al. 2005). However, most organisations are far from these leading practices. Relatively few organisations have applied IS automation to their end-to-end procurement systems. The top priority for Accounts Payable staff surveyed was to migrate off paper processes, (52% of respondents). Only 26% of companies surveyed had electronic invoicing capability (Ardent 2013).

### 2.1.3 Evidence: Metrics on performance and value – Business reports

In the business reports demonstrating leading practice, Best-in-Class enterprises were recognised for their outstanding performance. They attribute their reliance on key capabilities and technologies for their competitive advantages: 72% higher procurement contract compliance than all other companies; 52% higher rate of realized / implemented cost savings; and 37% higher rate of spend under management. Survey results also show these leading firms shared several common characteristics, including: 47% higher likelihood of instituting internal collaboration between key internal units; and 36% higher likelihood of leveraging an active spend analysis program (Aberdeen 2011).

"For 7 Eleven, strategic sourcing has translated into industry dominance. Over a two year period, they led all major rivals in same-Store Merchandise Sales Growth: 6.6% vs 3.5%. Merchandise Inventory Turns 38.3% vs 22.2%, and Merchandise Sales Per Employee $239 vs $98". 7-Eleven's sourcing strategy transformed the company by focusing on a small, strategic set of capabilities: in-store merchandising, pricing, ordering, and customer data analysis. The company reduced its capital assets and overhead while reducing head count 28% from 43,000 to 31,000 over 15 years and flattening its organizational structure, cutting managerial levels in half from 12 to six (Gottfredson et al. 2005).

In addition to the metrics utilised by Hackett (2014) presented on Table 1, that study found world-class Return on Investment (i.e., the ratio of cost savings on total purchases to total purchases) was at 900% return for investments in innovative procurement processes. At stage 5 on Table 1, the highest stage, the agenda for the procurement function is to support business objectives. The most appropriate metric for this stage is the business value generated for the corporation. By this stage, procurement has become a professional services provider that is a proactive, trusted agent of change. At this stage of maturity, a rich set of capabilities allows the company to harness the power of global supply markets for competitive advantage (Hackett 2014, p16).

## 2.2 Selected themes in literature on Information Systems (IS)

Information Systems literature has the potential to inform both academic and practitioner audiences on value co-creation by business networks as this field of study's core focus includes: business networks, business strategies, enterprise systems, inter-organisational systems, supply chain management, technology-enabled innovation and value creation. However, the application of these focus areas in the core business function of procurement / strategic sourcing across business networks for value co-creation has received limited attention. Several relevant sources were purposively selected (Miles and Huberman 1994) to identify and inform the current status on leading business practice in procurement.

### 2.2.1 Procurement transformation process - IS

IS-enabled business transformation represents a rich field for research and practice. An investigation of IT-enabled transformation of the supply chain (Swafford et al., 2008) identified a 'domino effect' among IT integration, supply chain agility and flexibility and competitive business performance. The necessity for future research in this area has been identified in several studies, with particular attention to: governance models, risk management tools, service portfolio management approaches and service bundling techniques, to better understand core determinants of competitiveness and success of service aggregators (e.g., Kohlborn et al. 2009). At a level beyond models, tools and techniques, a further call for research suggests organizational performance benefits from reframing supply chain management practice towards learning, networks, service-ecosystems and value co-creation (Lusch 2011).

| *Stage \** | *Role of procurement \** | *Value proposition\** | *Metrics in leading practice \** | *Metrics in IS theory* | *Metrics in Strategic Management theory* |
|---|---|---|---|---|---|
| 1. Supply Assurance | Buyer / planner | Right goods / services at right time & place | Availability; reduced inventory; growth in sales | Transaction cycle time, inventory costs | Operations process metrics, e.g., order cycle time |
| 2. Price | Negotiator | Right goods / services at right price | Reduced spend; avoided purchase costs | Transaction price & costs | Operations financial metrics, e.g., cost savings |
| 3. Total cost of ownership (TCO) | Supply expert, team leader & project manager | Shift from lowest price to TCO | Efficiency; contract compliance; cost savings; process automation | Total costs | Financial efficiency, e.g., return on assets |
| 4. Demand Management | Spend / budget consultant & relationship manager | Reduce demand activity, complexity & variability | Spend under management; higher-quality services; lower cost; spend visibility | Responsive-ness, user satisfaction | Demand response, e.g., create customer value |
| 5. Value Management | Trusted business advisor & change agent | Create business value from spend | Firm's goals & objectives delivered by business partnering | Enterprise effectiveness | Financial metrics, e.g., return on equity |

*Table 1. Stages in procurement: evolving roles, value propositions and metrics.*
\* Based on leading practices (Aberdeen 2011; Gottfredson et al. 2005; Hackett 2014). IS theory (Subramaniam and Shaw 2002). Strategy and Management theory (Ralston et al., 2015).

### 2.2.2 Characteristics of leading practitioners - IS

A survey of 38 major Australian organisations presents details of their current direct and indirect procurement practices. Those practices were then analysed together with the drivers and barriers of e-procurement. The study's results show that direct procurement was heavily dependent on traditional practices whilst indirect procurement was more likely to use "e" practices. Small-medium organisations were more nimble at adopting e-procurement practices. Technical issues dominate e-procurement barriers, with cost factors dominating e-procurement drivers (Hawking et al. 2004).

### 2.2.3 Evidence: Metrics on performance and value - IS

IS research papers describing metrics directly addressing world-leading procurement practices were not identified through literature searches, although a related call for such metrics to be developed was noted, "Performance measures and metrics need to be established for measuring the performance and suitability of IT in SCM" (Gunasekaran and Ngai 2004, p. 291) and endorsed (Akyuz and Erkan 2010).

To locate this paper's focus within the body of IS research literature, publications addressing related metrics are presented. A 2002 study of the value and impact of Web-based B2B procurement identified a series of metrics relevant to an earlier stage of technology transformation in procurement, from internal enterprise systems to external web-based systems. Procurement spend comprises two main categories. Direct purchases are used in the production of goods and services for customers. Indirect purchases are other expenses incurred by the business in order to operate and not specifically in the production of outcomes for sale to customers. The 2002 study examined indirect purchases primarily, so the resulting metrics derived from study of this cost category. Proceeding from operational measures to value management, the metrics utilised in that study were: transaction cycle time, inventory costs, transaction price and cost, total costs, responsiveness, (internal) user satisfaction and enterprise effectiveness in meeting its indirect purchase requirements (Subramaniam and Shaw 2002).

Further calls for action in addressing this issue critical to the IS field of study were identified in the IS literature. First, for IS researchers to venture beyond their traditional focus on developers, users, and managers and to consider other key stakeholders, including customers, employees, suppliers, stockholders, vendors, and governments. To provide practical benefit, IS success measures must capture all of these stakeholders. Second, to address the requirements of researchers and practitioners. Therefore, researchers and practitioners are encouraged to take on the challenge of developing IS measurement frameworks that reflect the ubiquitous impact of information systems from a personal level to a global level (Petter et al. 2012).

Recently, a comprehensive review of IS value literature identified three research gaps: to resolve the ambiguity of the 'IS business value' construct; to disaggregate IS from other investments; and, to analyse the processes by which IS can create business value. A research agenda proposed to address these gaps also seeks to investigate if strategic market advantage could be generated through integration of generic IS assets and complementary capabilities in a firm (Schryen 2013). This last proposal appears to be directly relevant to transformation of procurement.

## 2.3    Selected themes in literature on Strategy and Management

Noting the lack of relevant IS literature in addressing the transformation of the procurement field identified in industry reports, efforts were made to seek details from a related non-IS field of study. In 2013, a review was conducted for the premier publication, Journal of Supply Chain Management, into innovation in business networks from a supply chain perspective. This review of the current state of academic research and industry practice in this field highlighted the gap between practice and research. Noting "an obvious disconnect between the importance attributed by practice and the current state of academic research", it called for greater research efforts into innovations that actually span the supply chain network (Arlbjorn and Paulraj 2013, p3). Some explanation for this disconnect may be found in the conclusion of a paper previously published in the same journal, that resource-based theory suggests purchasing and supply chain management can rarely be sources of competitive advantage for firms (Ramsay 2001). To inform researchers about the gap noted above between theory and practice, a further range of academic sources was purposively selected. Although the number of sources relevant to the themes arising from business reports on the disruptive transformation of procurement functions was limited, several specific gaps in the literature were identified:

1. "There are only a few examples of how procurement firms classify the value co-creation potential of their suppliers (who provide services) and manage the transition of such suppliers into strategic partnerships."
2. "There are few examples in the literature of value co-creation of services. Much of the extant literature appears to focus on complex engineered products."
3. "There appears to be very little in the extant literature on the measurement of value co-creation opportunities or on supplier performance measurement systems that go beyond compliance to contractual agreements." (Nudurupati et al. 2015, p. 249).

### 2.3.1    Procurement transformation process - Strategy and Management

With few exceptions, research papers with some relevance to procurement transformation also tended to focus on specific aspects of the transformation, e.g., how a firm could mediate between a range of internal and external stakeholders to engage in a strategic partnership for value co-creation with multiple suppliers (Nudurupati et al. 2015) rather than attempting to describe the value co-creation process as a whole. A process for selecting strategic suppliers was identified. This entailed plotting prospective suppliers over a 2x2 matrix with the vertical axis Profit impact (Low to High) and the horizontal axis Supply risk (Low to High) (Kraljic 1983).

Some theoretical papers addressed issues within the themes identified in the business reports, e.g., "supply chain integration is viewed as a process by which a firm acquires, shares, and consolidates strategic knowledge and information internally throughout the firm and externally with supply chain partners" (Ralston et al., 2015) and, "in a value co-creation approach suppliers should focus on delivering benefits and provision of solutions in totality using a collectively exclusive collaborative partnership approach. Hence, the challenge of the purchaser company is to pursue an integrative and trans-disciplinary methodology" (Nudurupati et al. 2015, p. 251, based on Vargo and Lusch 2008; Bastl et al., 2012). It is, however, not clear that such research findings could inform business executives facing uncertainty about how to respond to drivers of business transformation. Or, should they be informed, if those research findings could be actionable.

### 2.3.2  Characteristics of leading practitioners - Strategy and Management

Unfortunately, and in contrast to practitioner literature, few characteristics of leading practitioners were addressed in the academic literature. Although multi-stakeholder collaboration and partnership have several advantages in sourcing decision-making, it was noted that its inappropriate implementation may result in failure, mainly due to a lack of appropriate commitment and involvement of the purchaser and suppliers. (Lock 1998).

### 2.3.3  Evidence: Metrics on performance and value - Strategy and Management

The theoretical papers included little attention to performance measures and value co-creation metrics. However, one paper focusing on strategic sourcing with multi-stakeholders through value co-creation identified metrics related to behaviours including: Always work in the best interest of the organisation; Be open; Solve first and settle later; and, Develop consistent approaches and the metrics: engagement, collaboration, cooperation and innovation (Nudurupati et al. 2015). It is not clear if these metrics relating to these behaviours could be generalizable to other companies seeking strategic sourcing opportunities such that they could be applied to distinguish between value creation by: world-class leaders and companies applying traditional procurement practices. However, a paper investigating the impact on firm performance from integration of strategic supply chain management across organisations identified: operations process metrics, e.g., order cycle time; operations financial metrics, e.g., cost savings; financial efficiency, e.g., return on assets; demand response, e.g., create customer value; and strategic financial metrics, e.g., return on equity, see Table 1 (Ralston et al. 2015).

## 3  Research design and method

As presented above, this paper has three integrated aims. First, to assist business executives to respond to the drivers of business transformation by investigating and describing how disruptive innovation is transforming Enterprise Supply Chain Management particularly the core procurement function. Second, to expedite IS researchers' efforts to increase the relevance of their work by examining digital disruption of procurement to close gaps between pioneering business practice and extant IS theory. Third, to increase the empirical impact of IS research by proposing rigorously-developed models of IS research and practice to be applied, tested and extended. The primary research questions derived from the three aims are:

a) Can research into disruptive innovation transforming procurement assist business executives responding to the drivers of this disruption? (Hawking et al. 2004; Kohlburn et al. 2004; Ralston et al. 2015).
b) Can research examining digital disruption of procurement assist IS researchers to increase the relevance of their work? (Agarwahl and Lucas 2005; Nudurupati et al. 2015).
c) Can IS research assist business executives by closing gaps between pioneering business practice and extant IS theory? (Gunasekaran and Ngai 2004; Schryen 2013).
d) Can IS research in this field propose models of IS theory and practice with empirical impact, e.g., in co-creating business value? (Lusch 2011; Petter et al., 2012).

Traditional academic approaches to problem solving in real-world contexts tend to follow the practices relevant to a particular discipline. However, in a complex, dynamic real-world situation the problem solving approach needs first to focus on the particular phenomenon (Gibbons et al., 1994). Innovative research and problem-solving approaches feature participative and interpretative techniques to examine an issue within context, e.g. grounded theory (Glaser and Strauss 1967), and induction (Babbie 2006). Inductive reasoning moves from the particular to the general, from a set of specific observations to the discovery of patterns (empirical generalizations) among all the given events (Babbie 2006).

Grounded theory (Glaser and Strauss 1967) is a method built on the key concepts: simultaneous collection and analysis of data, and theoretical sampling, where data to be collected is progressively determined by the theory under construction (Suddaby 2006). It is, "a methodology that is attentive to issues of interpretation and process and that does not bind one too closely to long-standing assumptions" (Suddaby 2006, p. 641). Grounded theory was conceived as an approach to theory development for situations where deductive approaches appeared inappropriate for the research aims. Observations and incidents are compared to determine their suitability to categories, categories and their properties are integrated, and the theory under development is scoped, articulated and presented in written form. Potential generalizability is proposed by the researchers. (Glaser and Strauss 1967).

This study's unit of analysis was the disruptive transformation of the procurement function from the perspectives of two organizations, a buyer and a supplier of business networking services, recognized as pioneers in the field. Data collected from reports and interviews represent social artifacts (Babbie 2006) produced by staff and executives interviewed who were participants in the companies' innovations. The most appropriate research approach for the study's aims and research questions was investigative field research. Induction, particularly suited for field studies (Babbie 2006) and grounded theory, suitable to generate knowledge in areas of human invention (Suddaby 2006) were selected. Both methods support development of transferable knowledge.

Consistent with the aims of this study, a rigorous field study (following Yin 2014) of these two world-class pioneering corporations, was undertaken to identify and to confirm: 1. gaps between current theory and activities reported as being leading practice (Aberdeen 2011; Hackett 2014) and, 2. possible gaps between reported leading practice and purposively selected (Miles and Huberman 1994) pioneers in disruptive practices.

The two global pioneers in procurement transformation were approached and agreed to participate in this study. Noting the general reluctance of business to disclose details that represent a source of market advantage, strict confidentiality was observed at all stages of the study, including in the initial submission of this publication. For the final submission the supplier approved the details for publication and naming. The buyer corporation, however, while approving the paper's content preferred anonymity at this stage and is referred to in this work as WBC. The supplier can be revealed as Ariba, a global of business networks for procurement. Participants in this study were: Vice-President Procurement WBC; Director Procurement Transformation WBC; Global Vice-President Ariba Value Engineering, and General Manager Ariba ANZ. Two are co-authors. Additional details were provided by other senior executives, including the President of Ariba.

Research rigour (Yin 2014) was achieved through preparation and implementation of a protocol for data collection and analysis across all sources. Reliability by the authors jointly collecting, analysing and categorising the data as well as determining and reviewing the emergent themes. Interview questions were jointly prepared and reviewed. Interview transcripts and advanced versions of the paper were reviewed by participants who granted final approval. Strategies to address research validity included: triangulation of data sources from corporate websites (externally and internally accessible), publicly available and internal corporate materials (reports, documents and presentations) and interviews with the executives responsible for the corporate responses between March and October 2015. The potential for participant bias was addressed through adherence to the formal processes of the research protocol and through triangulation of sources, participant review and final approval of research findings and outcomes by senior executives independent of the research team (Babbie 2006).

## 4    Field study

Enterprise Resource Planning (ERP) providers have been delivering Supplier Relationship Management solutions that include functionality for procurement and financial processes related to processing from requisition through to payment for over ten years. While the IS capabilities in theory supported many of the five stages in procurement's evolving value proposition (Table 1) there was limited improvement in performance value and metrics. In retrospect, it is clear that these IS capabilities had limitations, the most obvious been their focus mainly on direct costs. Direct costs are defined as cost of goods and services that are raw materials, components or sub-assemblies of the organisation's finished products.

The scope of procurement IS solutions varies between business reports. In this work, they are grouped into three major areas: functionality to support the requisition request to selection of goods or services process (sourcing); functionality that supports the purchase to payments process (P2P); and solutions to categorise and analyse organizational spend (spend analysis). Consistent with this study's aims, the research focus is on best practices in sourcing and related processes. Limited success in improving the key procurement transformation process metrics defined in section 2.1, Theme 2 (Aberdeen 2011) is confirmed by a more recent study of 245 business executives showing that 74% of respondent organisations still used manual processes in procurement (Ardent 2013, p7).

Second generation procurement IS combined external market changes and lessons learned from deployments of initial procurement systems, e.g., different spend categories required different management approaches and IS functionalities. Many 1$^{st}$ generation procurement solutions such as SAP SRM, were targeted at the 'Direct spend' category as manufacturing industries were early adopters of these technologies. The capabilities were not, however, optimized for the three other generic spend categories: 'Indirect spend', the overheads and other corporate expenses required to operate but not directly constituent in finished goods and services; 'Travel & Expenses' incurred by employees outside of remuneration; and 'Complex Services' including contract and contingent labour. These later spend categories have become an increasingly important components of overall spend for business, government and societal organizations in recent years (General Manager Ariba ANZ).

Drivers of this change included the Global Financial Crisis (GFC) of 2008-9. "Many organizations responded to the GFC impacts of reducing revenue and delayed customer spending by reducing overall headcount. As the organization still needed to perform many of the same functions, organizations expanded contract and contingent services to compensate. When the market crisis passed, many organizations realized that higher proportions of contract and contingent labour provided higher organizational agility and as a result in many areas permanent headcount never recovered to post GFC levels" (Global Vice President Ariba Value Engineering).

Current 2$^{nd}$ generation procurement IS solutions address these broader requirements by providing innovative new approaches, of which the most disruptive is arguably the business network. "The business network solves an increasingly important problem for organisations, how to evolve their SCM strategy from competition to collaboration and innovation" (Global Vice President Ariba Value Engineering). The Ariba network is an early example of such an IS-based business network. Globally, it facilitates about $650 billion in transaction value annually, incorporates over 1.5 million suppliers and extends innovative new procurement processes across the supply ecosystem (President of Ariba, Ariba Live Opening Keynote 2015, Las Vegas).

An early innovator with Ariba's business network, WBC is a leading entertainment industry organization. It is one of the world's most diversified casino-entertainment companies. Since its beginning in Nevada, 60 years ago, WBC has grown through development and now operates casinos on four continents. It has more than 10 casinos with over one million square feet of gaming space, 20,000 hotel rooms and 50,000 employees. WBC is committed to being a positive corporate citizen, contributing to improving economic, social and environmental qualities of life wherever it operates. In procurement, this means there is a goal to engage with local and diverse owner-suppliers in each operating location.

WBC is a leader in its field and is experiencing change pressures due to external factors including; the need to be global to maintain competitive scale, new alternate business models including online gaming, margin and pressure and constant scrutiny in what is a highly regulated industry, and market saturation is some geographic locations. WBC sought to drive improvement in certain financial metrics in 2010. Analysis quickly identified that investment in and transformation of procurement related activities could drive some of the business benefits and improvements sought. One of the major financial metrics examined is working capital. The approach to and levels of working capital in any organization have a profound effect of the operating strategies of the business. Days Sales Outstanding (DSO) and Days Payable Outstanding (DPO) are key metrics that drive working capital. Due to the high cost of capital in the business, WBC elected to focus on sourcing, contracting, invoicing and payment holistically. Their key objectives were to drive cost out of supply chain, better understand spend, boost compliance and proactively manage spending (Dir. Procurement Transformation WBC).

"WBC took the decision to further invest in best-in-market procurement skills. With the introduction of new procurement skills, the organization embarked on a journey in pursuit of best practices that led to implementation of an IS-based business network as a core tenant of that best practice and change" (Vice-President Procurement WBC). In 2010 WBC had few specialist spend category managers and had a largely decentralized structure for sourcing in all spend areas except direct spend, with individual departments having control over individual buyers. In terms of procurement evolution (see Table 1) WBC was poised to be a stage 5 procurement organisation.

An early decision was to adopt the Ariba's business network. An example of IS driven innovation provided by the business network is 'Supplier Discovery'. This enables buyers to quickly poll the more than 1.5 million active network suppliers to locate potential new sources of supply, in completely new categories where the individual buyer may have no previous experience or expertise. As a result of the Discovery capability, a buyer can not only significantly reduce the time to deliver a short list of suppliers to the business that meet specific criteria, but can also for the first time, easily represent WBC's socially oriented business objectives within the sourcing process, e.g., the buyer can easily identify and/or specify supplier-required criteria such as minority owned, sustainable or local business. The results to date from the WBC's procurement transformation are significant: More than 72% of overall spend under management; 9% savings in overall cost on multi-million dollar deals; 45% savings in time to find a low-cost local supplier; 90% reduction in sourcing request turnaround time; and 50,000 invoices processed per Accounts payable FTE versus 2010 internal benchmark of 15,000 (Dun & Bradstreet 2015).

## 5　Research findings

The research approach, induction and application of Grounded Theory (Glaser and Strauss 1967) in conjunction with Yin's (2014) guidelines for rigorous qualitative research, successfully supported the development of generalizable outcomes from this work. New IS capabilities proved to be critical in WBC's evolution of Procurement Sourcing best practices. The focus on matching buyer needs with the complete ecosystem of supplier options, takes the category knowledge of a professional buyer and codifies this capability making new forms of supplier buyer interaction possible. Specifically, it enables buyers to focus at the high end of the stages of procurement model (Table 1). While Table 1 assists researchers and executives to identify and evaluate each stage in the procurement maturity process, it does not provide guidance on how to progress. Based on analysis of the research findings a generic process, the Business Network Value Creation (BNVC) Model, is proposed to assist executives accelerate the rate of maturity in their procurement function (Figure 1).

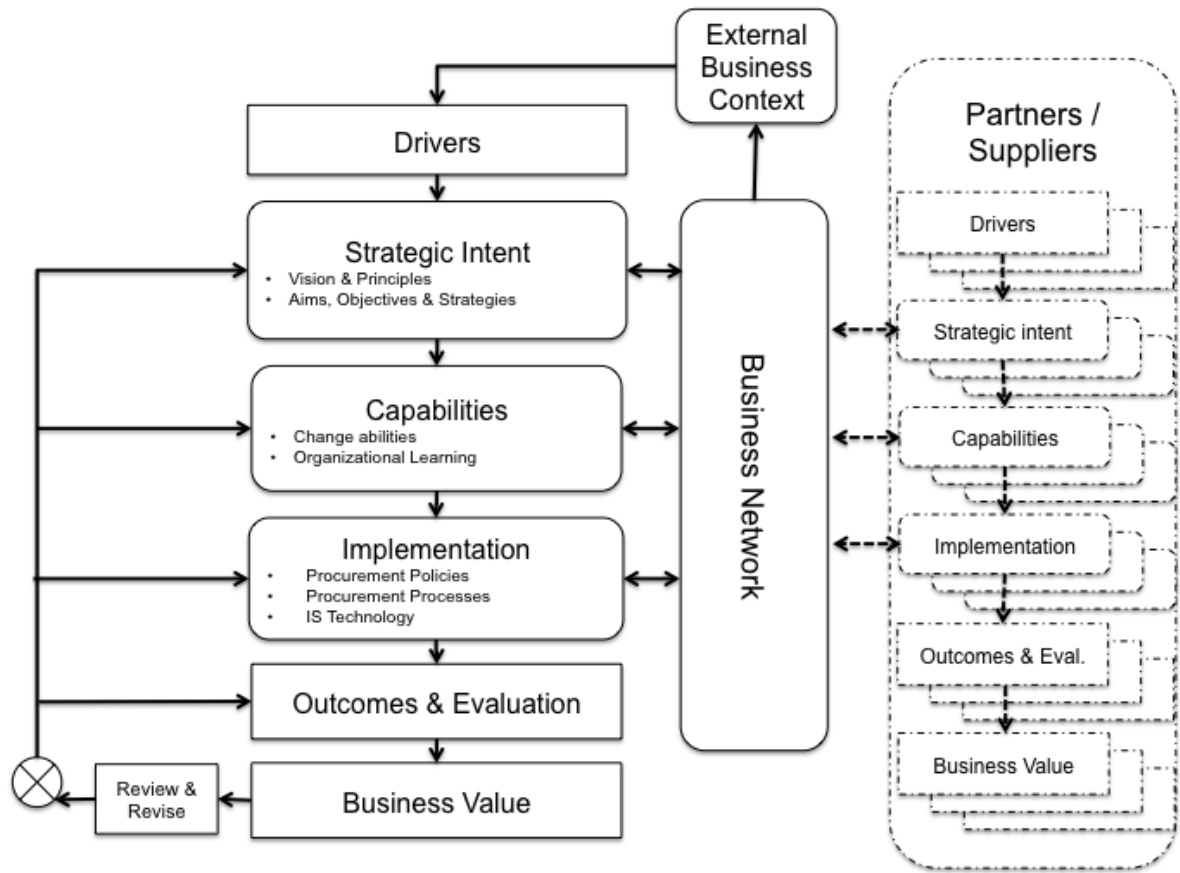

*Figure 1. Business Network Value Creation (BNVC) model: Buyer–Supplier integration of critical processes to generate value through the disruptive innovation of business networks (source: field study)*

The BNVC model acknowledges that transforming the maturation process is not confined to the 'hub' organizations' boundaries, but incorporates the innovation abilities of the suppliers. In effect, transformation occurs simultaneously in multiple processes with internal processes running vertically, and supply chain collaborative processes, horizontally across organisations. The BNVC model seeks to address previously identified issues with procurement processes (Gunasekaran and Ngai 2004; Schryen 2013). It particularly addresses the key issue of linking procurement to, and aligning with, business objectives. Critically, the business network is presented as core to an organizations' procurement maturation process for its two essential roles: 1. automating process flows across the supply chain that provides visibility and control to both buyers and suppliers; and 2. enabling agility, underpinning both the speed of procurement maturation and its capacity for innovation.

# 6   Discussion

This work's aims and research questions sought to assist business practitioners and IS researchers to unite in overcoming their currently separate challenges. For business executives, this is to learn how acknowledged world-leading buyers and suppliers have been able to generate business value from procurement-oriented business networks. For IS researchers it was to demonstrate the potential relevance of their work to practitioners and academic colleagues. The aims and research questions have been realised through: a selected review of practitioner reports that identified disruptive innovations in procurement processes and evaluation metrics; further reviews of relevant literature in Information Systems that revealed the lack of IS research in this field and also in the Strategy and Management fields; and a field study of world-class pioneering practice demonstrating the current state of leading practice.

The study's outcomes are: Table 1, illustrating stages of development in procurement practices with perspectives on the changing role of procurement value propositions and metrics for evaluating developments from theory and practice; a field study presenting previously unavailable insights into leading practices in buyer-supplier relationships; and conception of the BNVC model of how executives could implement, and researchers investigate, processes to create value through the core business function, procurement (Figure 1). This study is limited in its sample of two companies in a single buyer-seller business function. However, and despite this limitation, this study reveals fresh insights into business transformation of a core business function. IS researchers seeking to improve their relevance will benefit from the findings and contributions of this work. Further efforts will explore this limitation.

# 7   Conclusion

In a dynamically changing world, businesses must transform to survive. However, business transformation is complex, high risk and uncertain. Through a rigorous field study of two world-class pioneering corporations, a buyer and a supplier, this paper explores how disruption is transforming Enterprise Supply Chains. Lessons, contributions and their implications for current IS theory and practice are discussed. This work's contributions highlight disconnects between academic research and leading business practice. A comparison of reports on business practice describing processes and the means of evaluating those processes demonstrates the disruptive nature of current leading practice and the gaps between research and practice. The BNVC model (Figure 1) illustrates how and where business executives could learn from these experiences and apply those lessons within their own contexts.

This work also enables IS researchers to respond to calls on the IS field of study to increase the relevance of its research and learning (Agarwal and Lucas 2005). Design of IS solutions to facilitate transformation of core business functions has broad international relevance that falls within the core of the IS field. The previously unavailable insights into the procurement practices of pioneering companies also show how research contributions based on IS practices can contribute directly to reducing business uncertainties through development and application of theory based on practice. Since business transformation is dynamic, the BNVC model and details of procurement transformation will need to be revised and expanded by future studies. The relevance of research in this area would appear assured.

## Acknowledgements

Grateful appreciation is due to the executives and staff of Ariba and 'WBC' for sharing their revelatory experiences, from all who seek every opportunity to learn.

## Copyright